\documentclass{elsart}
\usepackage[dvips]{graphics,graphicx,color}
\usepackage{indentfirst,latexsym,amsfonts,amssymb,amsmath}

\begin{document}

\begin{frontmatter}

\title{Long-term damage induced by hadrons in silicon detectors for uses at the 
LHC-accelerator and in space missions}
\author[univ]{I. Lazanu} and
\author[iftm]{S. Lazanu\thanksref{cor}},

\address[iftm]{National Institute for Materials Physics, POBox MG-7, Bucharest-Magurele, 
Romania}
\address[univ]{University of Bucharest, Department of Nuclear Physics, POBox MG-11, 
Bucharest-Magurele, Romania}

\thanks[cor]{corresponding author: fax: +40-21-4930267, e-mail:lazanu@alpha1.infim.ro}
\begin{abstract}
In the present paper, the phenomenological model developed by the authors in previous 
papers has been used to evaluate the degradation induced by hadron irradiation at the 
future accelerator facilities or by cosmic protons in high resistivity silicon detectors. 
The damage has been analysed at the microscopic (defects production and their evolution 
toward equilibrium) and at the macroscopic level (changes in the leakage current of the 
p-n junction). The rates of production of primary defects, as well as their evolution 
toward equilibrium have been evaluated considering explicitly the type of the projectile 
particle and its energy. Vacancy-interstitial annihilation, interstitial migration to 
sink, complex ($VP$, $VO$, $V_2O$, $C_iO_i$ and $C_iC_s$) and divacancy formation are 
taken into account for different initial silicon material. The influence of these defects 
on the leakage detector current has been calculated in the frame of the Schokley-Read-Hall 
model. 

\medskip
\textbf{PACS}: \\
29: Experimental methods and instrumentation for elementary-particle and nuclear physics\\
81: Materials science\\
78: Optical properties, condensed-matter spectroscopy and other interactions of radiation 
and particles with condensed matter

\end{abstract}

\end{frontmatter}

\section{Introduction}

The use of silicon detectors at the new generation of accelerators or in space experiments 
poses severe problems due to changes in the properties of the material after long time 
irradiation, and consequently influences the performances of detectors.

The phenomenological model developed by the authors in previous papers has been used to 
evaluate the degradation induced by hadron irradiation at the future LHC accelerator 
facilities and by cosmic protons in high resistivity silicon detectors, in two types of 
silicon: "standard" material (10$^{14}$ cm$^{-3}$ atoms of phosphorus, $2\cdot 10^{15}$ 
cm$^{-3}$ atoms of oxygen, and $5\cdot 10^{15}$ cm$^{-3}$ atoms of carbon), and "oxygened" 
one (containing  10$^{14}$ cm$^{-3}$ atoms of phosphorus, $4\cdot 10^{17}$ cm$^{-3}$ atoms 
of oxygen, and $5\cdot 10^{15}$ cm$^{-3}$ atoms of carbon) respectively. The damage has 
been analysed at the microscopic (defect production and their evolution toward 
equilibrium) and at the macroscopic level (changes in the leakage current of the p-n 
junction).

These theoretical estimates permit to draw conclusions about the damages induced in 
silicon in various irradiation conditions and for different semiconductor materials.

\section{Main hypothesis of the model}

In the model, the effects of irradiation conditions and various initial impurities in the 
starting material are discussed in the a quantitative manner that the defect production 
and their evolution toward stable defects during and after irradiation in silicon is 
calculated. The model supposes three steps. 

In the first step, the incident particle, having kinetic energy in the intermediate up to 
high energy range, interacts with the semiconductor material. The peculiarities of the 
interaction mechanisms are explicitly considered for each kinetic energy 
\cite{nim388,nim432}. 

In the second step, the recoil nuclei resulting from these interactions lose their energy 
in the lattice. Their energy partition between displacements and ionisation is considered 
in accord with the Lindhard theory (see reference \cite{lind} and authors' contributions 
\cite{lindnoi}).

A point defect in a crystal is an entity that causes an interruption in the lattice 
periodicity. In this paper, the terminology and definitions in agreement with M. Lannoo 
and J. Bourgoin \cite{bourg} are used in relation to defects.

We denote the displacement defects, vacancies and interstitials, as primary point defects, 
prior to any further rearrangement. After this step the concentration of primary defects 
is calculated.

The concentration of the primary radiation induced defects per unit fluence ($CPD$) in 
silicon has been calculated as the sum of the concentration of defects resulting from all 
interaction processes, and all characteristic mechanisms corresponding to each interaction 
process, using the explicit formula (see details, e.g. in references \cite{np61b,nim413}):
\begin{equation}
CPD	\left(E\right)= \frac{N_{Si}}{2E_{Si}} \int \sum _{i} \left( \frac{d\sigma}{d\Omega} 
\right)_{i,Si} L(E_{Ri})_{Si} d\Omega=\frac{1}{N_A} \frac{N_{Si}A_{Si}}{2E_{Si}} 
NIEL\left(E\right)
\end{equation}
where $E$ is the kinetic energy of the incident particle, $N_{Si}$ is the atomic density 
in silicon, $A_{Si}$ is the silicon atomic number, $E_{Si}$ - the average threshold energy 
for displacements in the semiconductor, $E_{Ri}$  - the recoil energy of the residual 
nucleus produced in interaction $i$, $L(E_{Ri})$ - the Lindhard factor that describes the 
partition of the recoil energy between ionisation and displacements and 
$(d\sigma/d\Omega)_i$ - the differential cross section of the interaction between the 
incident particle and the nucleus of the lattice for the process or mechanism $i$, 
responsible in defect production. $N_A$ is Avogadro's number. The formula gives also the 
relation with the non ionising energy loss ($NIEL$), the rate of energy loss by 
displacement $(dE/dx)_{ni}$ \cite{vangineken,burketoti}. 

The kinetic energy dependence of CPD for pions (from reference \cite{nim419}) and for 
protons \cite{vangineken,burke} is used in the present calculations. $CPD$ versus the 
kinetic energy of pions and protons is presented, e.g., in Figure 1 of reference 
\cite{physscr}.

The basic assumption of the model is that primary defects, vacancies and interstitials, 
are produced in equal quantities and are uniformly distributed in the material bulk. They 
are produced by the incoming particle, or thermally - only Frenkel pairs are considered. 
The generation rate of primary defects is a sum of two components:
\begin{equation}
G=G_R+G_T
\end{equation}
where $G_R$ accounts for the generation by irradiation, and is calculated as:
\begin{equation}
G_R=CPD(E)\times \Phi(E)
\end{equation}
with $\Phi(E)$  the flux of considered incident particles, and $G_T$  for thermal 
generation.

In silicon, vacancies and interstitials are essentially unstable and interact via 
migration, recombination and annihilation or produce other defects. The concentration of 
primary defects represents the starting point for the following step of the model, the 
consideration of the annealing processes, treated in the frame of the chemical rate 
theory. A review of previous works about the problem of the annealing of radiation induced 
defects in silicon can be found, e.g. in Reference \cite{nimb183}.

After silicon irradiation, the following stable defects have been identified 
experimentally and confirmed: $Si$, $VP$, $VO$, $V_2$, $V_2O$, $C_iO_i$, $C_i$, $C_iC_s$ 
(after the compilations from references \cite{bourg,mara}).
Vacancy-interstitial annihilation, interstitial migration to sinks, divacancy and vacancy 
- impurity complex formation ($VP$, $VO$, $V_2O$, $C_i$, $C_iO_i$ and $C_iC_s$) have been 
considered supposing the following chemical reaction scheme:
\begin{equation}                                                        
V+I\ _{\overleftarrow{G}} ^{\underrightarrow{K_1}}\text{annihilation}
\end{equation}
\begin{equation}                                                       
I\stackrel{K_2}{\rightarrow } \text{sinks}
\end{equation}
\begin{equation}							
V+O\ _{\overleftarrow{K_4}} ^{\underrightarrow{K_3}}\ VO
\end{equation}
$VO$ is the $A$ centre.
\begin{equation}							
V+P\ _{\overleftarrow{K_5}} ^{\underrightarrow{K_3}}\ VP
\end{equation}
$VP$ is the $E$ centre.
\begin{equation}						
V+V\ _{\overleftarrow{K_6}} ^{\underrightarrow{K_3}}\ V_2
\end{equation}
\begin{equation}						
V+VO\ _{\overleftarrow{K_{7}}} ^{\underrightarrow{K_{3}}}\ V_2O
\end{equation}
\begin{equation}
I+C_s\stackrel{K_1}{\rightarrow }  C_i
\end{equation}
\begin{equation}
C_i+O_i\stackrel{K_8}{\rightarrow } C_iO_i
\end{equation}
\begin{equation}
A+I\stackrel{K_9}{\rightarrow }  O
\end{equation}
\begin{equation}
C_i+C_s\stackrel{K_8}{\rightarrow }  C_iC_s
\end{equation}

$K_i$ with  i = 1, $3 \div 8$ are the reaction constants. Some considerations about the 
determination of the reaction constants are given in references \cite{physscr,nimb183} and 
their concrete values are given in reference \cite{arXiv}, and \cite{fl2002}.
Without free parameters, this model is able to predict the absolute values of the 
concentrations of defects and their time evolution toward stable defects, starting from 
the primary incident particle characterised by type and kinetic energy.

\section{Radiation environment at the LHC accelerator and in the near Earth orbits}
In the present paper, two types of applications of silicon detectors are emphasised: for 
the tracker of experiments at the LHC accelerator, and for space missions in the near 
Earth orbit, as, for example experiments at the International Space Station.

At the luminosity of $10^{34}$ cm$^{-2}$s$^{-1}$, and assuming an inelastic 
non-diffractive cross section of 80 mb, the LHC will produce on average $8\cdot 10^8$ 
inelastic p-p events per second, creating an extremely hostile radiation environment. For 
radiation studies, the bunch structure of LHC is not significant. The only scaling 
parameter for doses and fluences is the inelastic interaction rate.

The central tracker is expected to be exposed to a primary particle flux from the 
interaction region, and the main concern is radiation damage of silicon detectors. Without 
loss of generality, the radiation field simulated for the CMS silicon tracker geometry is 
considered in the following calculations \cite{cms}. The high magnetic field imposes a   
$p_T$ cut-off on charged particles, so that a significant proportion of the most damaging 
low energy particles never reach the outer tracker layers. In addition, the average 
kinetic energy rises with increasing pseudorapidity. The spectra of charged hadrons 
(pions, kaons and protons) simulated for the positions of the silicon layers are taken 
from Reference \cite{cms}. The flux decreases by a factor 50  going from $r=20$ cm  to 
$r=100$ cm, and most of the low energy particles disappear. The hadrons are predominantly 
low-energy charged pions and protons, that are present in different amounts and have 
different energy spectra as a function of the distance in respect to the interaction 
point, and of the pseudorapidity. In all cases, the pions are the dominant particles.

In figure 1 the energetic differential generation rate of defects is presented. Each 
spectrum has been obtained as a convolution of the simulated hadron flux in the tracker 
cavity for CMS experiment and the energetic dependence of concentration of primary defects 
($CPD$) for protons and pions in the same energy range.
\begin{figure}[ht]
\centering
\includegraphics[width=.8\textwidth, clip, angle=90]{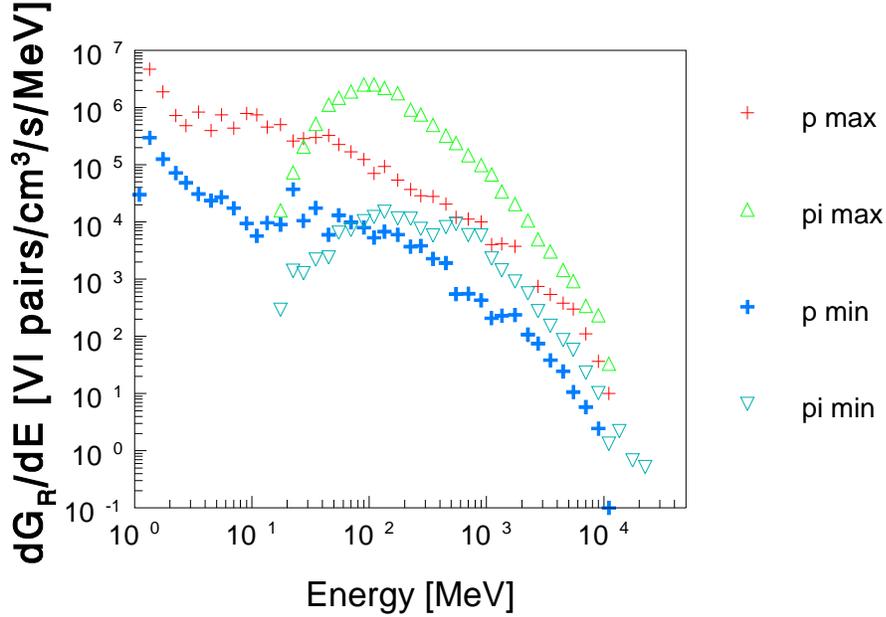}
\caption{\small{Differential energetic generation of vacancy-interstitial pairs for pions 
and protons, for two positions (1): $r=20$ cm, $z=0\div 140$ cm, and (2): $r=100$ cm, 
$z=140\div 280$ cm. in the tracking cavity of LHC.}}
\label{f1}
\end{figure}

The calculations have been performed for two extreme positions in the tracker cavity: (1):   
$r=20$ cm, $z=0\div 140$ cm, and (2): $r=100$ cm, $z=140\div 280$ cm. 

In the figure, the area under each curve represents the integral generation rate of 
vacancy-interstitials pairs. The values obtained for the first position (1) are: 
$6.2\cdot10^8$ VI pairs/cm$^3$/s  for pions and $5.6\cdot 10^7$ VI pairs/cm$^3$/s  for 
protons, while for the second one (2) these are $8.1\cdot 10^6$ VI pairs/cm$^3$/s  and   
$3.1\cdot10^6$ VI pairs/cm$^3$/s for pions and protons respectively.

We would like to mention that the main contribution for protons comes from the lowest 
energy region, while for pions the maximum shifts from around 200 MeV  to 800 MeV passing 
from position (1) to position (2). For pions, the  $CPD$ dependence is cut at 20 MeV. 

The second type of application discussed in the present paper refers to the radiation 
field produced by cosmic rays. From these particles, the most important contribution comes 
from protons. The primary proton spectrum in the kinetic range  0.2 to 200 GeV, in the 
neighbourhood of the Earth, at an altitude of about 380 km, was measured by the Alpha 
Magnetic Spectrometer (AMS) during space shuttle flight STS-91. The complete data set 
combining three shuttle altitudes and including all known systematic effects is given in 
reference \cite{ams}. The convolution of this spectrum with the CPD for protons is 
presented in Figure 2, and it corresponds to a generation rate of vacancy-interstitial 
pairs of $2\cdot 10^2$ VI pairs/cm$^3$/s. This represents a nearly seven order of 
magnitude lower generation rate in respect to the higher one calculated for the tracking 
cavity of the LHC experiments.
\begin{figure}[ht]
\centering
\includegraphics[width=.8\textwidth, clip, angle=90]{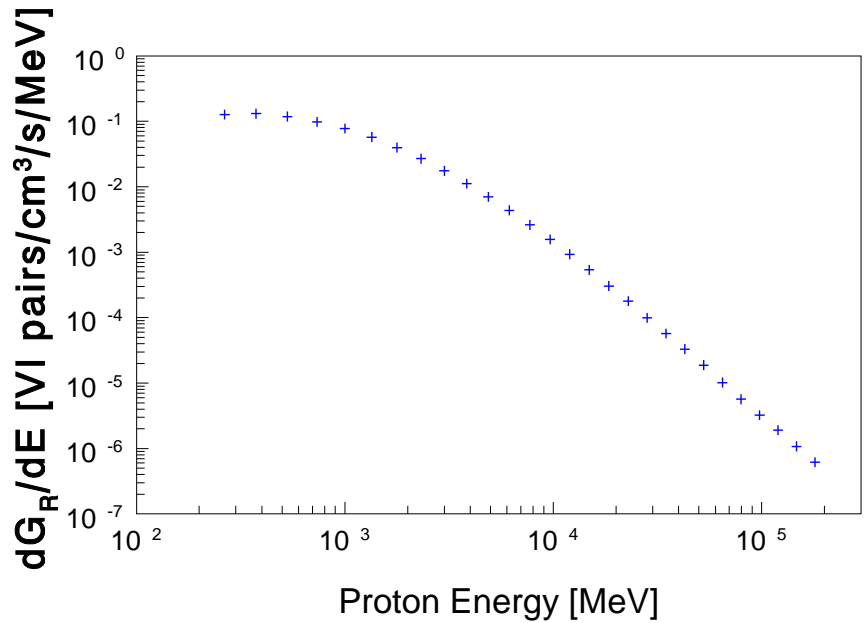}
\caption{\small{Differential energetic generation rate of vacancy-interstitial pairs by 
cosmic protons in the space near the Earth.}}
\label{f2}
\end{figure}

\section{Estimated damage in silicon detectors}

Silicon used in high energy physics detectors is n-type high resistivity ($1\div6$ 
K$\Omega\cdot$ cm) phosphorus doped FZ material.
In the last decade a lot of studies have been performed to investigate the influence of 
different impurities, especially oxygen and carbon, as possible ways to enhance the 
radiation hardness of silicon for detectors in the future generation of experiments in 
high energy physics - see, e.g. references \cite{moll,mcevoy}. Some peoples consider that 
these impurities added to the silicon bulk modify the formation of electrically active 
defects, thus controlling the macroscopic device parameters. The effect of oxygen in 
irradiated silicon has been a subject of intensive studies in remote past. Empirically, it 
is considered that if the silicon is enriched in oxygen, the capture of radiation 
generated vacancies is favoured by the production of pseudo-acceptor complex 
vacancy-oxygen. Interstitial oxygen acts as a sink for vacancies, thus reducing the 
probability of formation of divacancy related complexes, associated with deeper levels 
inside the gap. These conclusions are confirmed by the present model.

The concentrations of interstitial oxygen and substitutional carbon are silicon are 
strongly dependent on the growth technique. In high purity Float Zone Si, oxygen 
interstitial concentrations are around 10$^{15}$ cm$^{-3}$, while in the oxygenation 
technique developed at BNL, an interstitial oxygen concentration of the order $5\cdot 
10^{17}$ cm$^{-3}$ is obtained. These materials can be enriched in substitutional carbon 
up to $[C_i]\approx 1.8\cdot 10^{16}$ cm$^{-3}$.

\subsection{Changes in the microscopic material properties}
In Figure 3 a) to f), the formation and time evolution of the vacancy-oxygen, 
vacancy-phosphorus, divacancy, divacancy-oxygen, carbon interstitial-oxygen interstitial 
and carbon interstitial-carbon substitutional are presented. Two types of silicon have 
been considered: the "standard" material, containing the following impurities 
concentrations: $10^{14}$ cm$^{-3}$  atoms of phosphorus, $2\cdot 10^{15}$ cm$^{-3}$  
atoms of oxygen, and  $5\cdot 10^{15}$ cm$^{-3}$ atoms of carbon; and the "oxygened" one, 
containing  $10^{14}$ cm$^{-3}$  atoms of phosphorus,  $4\cdot 10^{17}$ cm$^{-3}$  atoms 
of oxygen, and  $5\cdot 10^{15}$ cm$^{-3}$  atoms of carbon respectively, and for the two 
generation rates of $VI$ - pairs discussed above, namely $7\cdot 10^8$  and $2\cdot 10^2$ 
$VI$ pairs/cm$^3$/s. The curves in the figures are labelled as follows: (1): standard 
silicon, LHC irradiation rate, (2): oxygenated silicon, LHC irradiation; (3): standard 
silicon, cosmic irradiation rate; and (4): oxygenated silicon, cosmic exposure. 
\begin{figure}[ht]
\centering
\includegraphics[width=.9\textwidth]{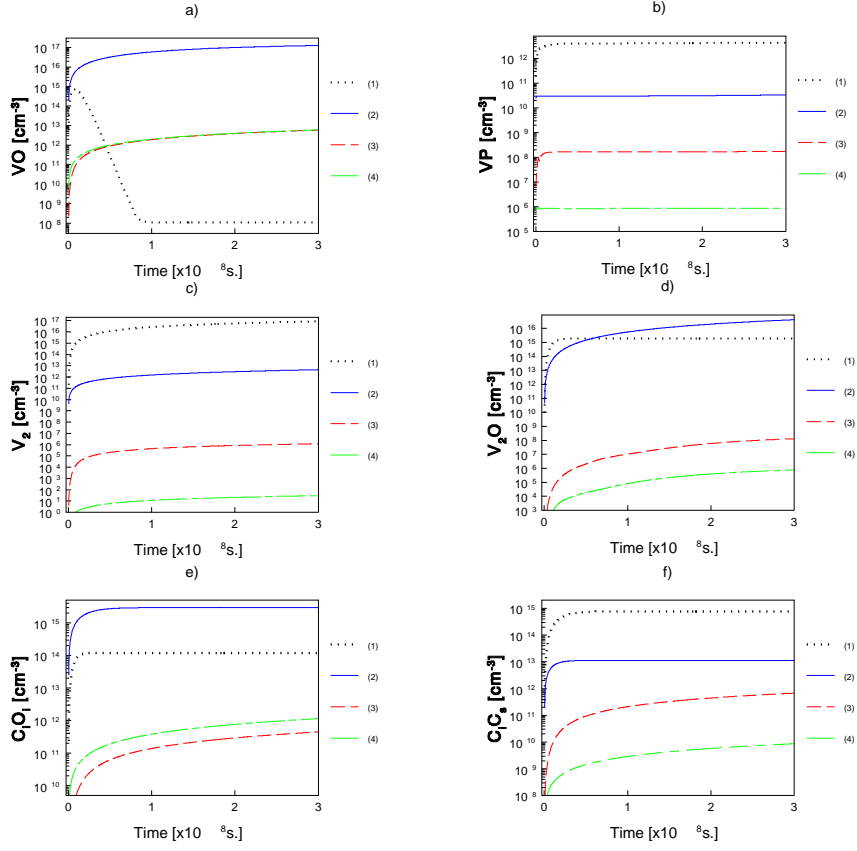}
\caption{\small{Time dependence of the concentrations of a): $VO$, b): $VP$, c): $V_2$, 
d): $V_2O$, e): $C_iO_i$ and f): $C_iC_s$ induced in standard and oxygenated silicon, in 
conditions of LHC and cosmic continuous irradiation rates (see text).}}
\label{f3}
\end{figure}
As could be observed from the figure, the content of oxygen in silicon influences 
especially defects formation in the case of high rates of generation of 
vacancy-interstitial pairs. The increase of the initial oxygen concentration in silicon, 
conduces, after ten years of operation in the LHC environment, characterised by a high and 
constant generation rate, to the increase of the concentrations of $VO$ and $C_iO_i$ 
centres, and to the decrease of the concentrations of $V_2$, $VP$ and $C_iC_s$ ones. With 
the increase of oxygen concentration, an increase of the $V_2O$ generation rate is 
observed. It is interesting to observe that in almost all cases, an equilibrium in reached 
between generation and annealing, and a plateau is obtained in the time dependence of the 
concentrations. The slowest is, in this respect, $V_2O$, that has the highest binding 
energy.

As underlined above, vacancy-oxygen formation in oxygen, enriched silicon is favoured in 
respect to the generation of $V_2$, $V_2O$ and $VP$ centres. At high oxygen 
concentrations, the concentrations of $VO$ centres attain a plateau during the 10 years 
period considered.

After cosmic proton irradiation, the effects are strongly different. For this generation 
rate, the increase of the oxygen concentration produces the decrease of the concentration 
of all centres, with the exception of the $VO$ concentration, that, at these rates, is not 
influenced by the oxygen content, and of the $C_iO_i$ concentration, where an increase is 
observed. As a consequence of the small rate of generation of vacancy-interstitial pairs, 
after ten years of operation, the equilibrium between generation and annealing is not 
reached, the concentrations of defects being, with the exception of $VP$ (that has a 
relatively low binding energy), slightly increasing functions of time.

All the processes have been calculated for 20$^o$C temperature. Thermal generation has 
been taken into account in all cases, although it is important only for the silicon 
exposed to cosmic protons.

Also, it is necessary to emphasise the importance of the irradiation and annealing 
conditions (initial material parameters, type of irradiation particles, energetic incident 
particle spectra and their flux, temperature) on defect evolution. These aspects have been 
discussed in previous papers \cite{nimb183,fl2002}.
 
\subsection{Macroscopic modifications of detector parameters - the leakage current}
The dark current of a reverse biased $p-n$  junction is composed of the following terms: 
the drift current, due to the drift of minority carriers, the generation current, due to 
carrier generation on the midgap energy levels inside the depleted region and surface and 
perimetral currents, dependent on the environmental conditions of the surface and the 
perimeter of the diode. The formation, during and after irradiation, of defects with 
associated energy levels inside the gap conduces to the increase of the generation 
current, since the ease with which a mobile carrier can traverse the gap is greatly 
enhanced by intermediate levels.

Inside the depleted zone, $n,p\ll n_t$  ($n_t$ is the intrinsic free carrier 
concentration); each defect with a bulk concentration $N_T$  causes a generation current 
per unit volume of the form \cite{borchi}:
\begin{equation}
I=qU=q<v_t>n_i\frac{\sigma _n \sigma _pN_T}{\sigma _n \gamma _n e^{(E_T-E_i)/k_BT}+\sigma 
_p \gamma _p e^{(E_i-E_T)/k_BT}}
\end{equation}
where $\gamma _n$  and $\gamma _p$  are degeneration factors, $\sigma _n$ ($\sigma _p$)  
are the cross sections for majority (minority) carriers of the trap, $E_i=(E_c-E_v)/2$   
and $<v_t>$  is the average between electron and hole thermal velocities.
In the Shockley-Read-Hall model used for the calculation of the reverse current, each 
defect has one level in the gap, and the defect levels are uncoupled, thus the current is 
simply the sum of the contributions of different defects. 
Two parameters characterising the defects enter in the calculation of the generation 
current: their energy position in respect to the intrinsic level, and their cross section. 
Only near midgap energy levels are important, and in this paper the contributions coming 
from $V_2O$, $V_2$ and $VP$ have been taken into account. An average between the values of 
the energy levels and cross sections reported in the literature (see compilations 
\cite{bourg,mara}) for $V_2$ and $VP$ have been introduced in the calculations, and 
averaged while for $V_2O$ this is true only for the energy level. In the lack of reported 
data for the cross section, for the $V_2O$  centre, the value $10^{-16}$ cm$^{-2}$ has 
been used.

Divacancy has three energy levels in the band gap. The equilibrium statistics for this 
case is formally different for this case from that for the same number of independent 
levels, since the occupancy of the levels is now interdependent. In the present 
calculations, the interaction of the $V_2$ energy levels has been neglected (in accord 
with Sah and Schotkley affirmation \cite{SRH} that the independent and interacting energy 
level cases are indistinguishable if the energy levels are more than a few $k_BT$ apart).

In Figure 4, the separate contributions of $V_2$O, $V_2$ and $VP$ defects to leakage 
current for "standard" and "oxygenated" silicon, in the irradiation conditions supposed 
for LHC and in the cosmic near Earth orbits, are represented. The calculations have been 
performed for 20$^o$C, under continuous irradiation. The bands representing the maximal 
uncertainties due to the energy position of the defects and to their cross sections are 
also drawn as dotted lines.
\begin{figure}[ht]
\centering
\includegraphics[width=.95\textwidth]{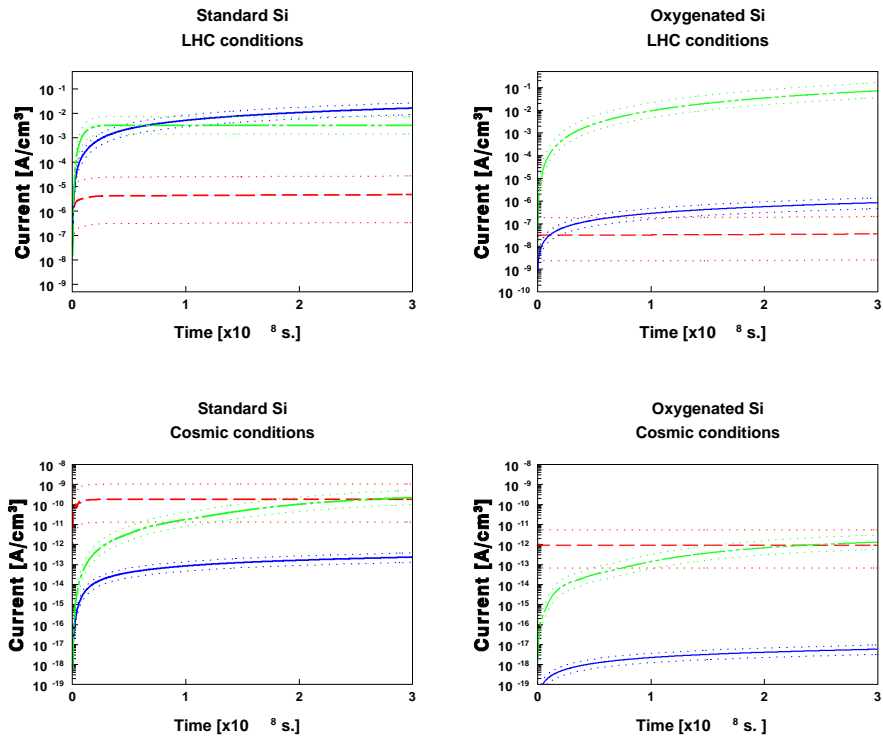}
\caption{\small{Time dependence of the generation current due to divacancy (continuous 
line), $VP$ (dashed line) and $V_2O$ (dashed dotted line).}}
\label{f2}
\end{figure}
On can observe that in the conditions of LHC generation rate, the values of the total 
current are nearly the same for standard and oxygenated silicon: the higher contributions 
from the $V_2$ and $VP$ centres (standard silicon) are counterbalanced by the increase of 
the $V_2O$ concentration (oxygenated silicon), that is the nearest to the midgap. 
For smaller generation rates, oxygenated silicon is a better choice from the point of view 
of the leakage current, as could be seen from the case of cosmic irradiation.

\section{Summary}
The phenomenological model developed previously to explain defect generation and evolution 
in silicon has been used to evaluate the damage induced by hadron fields, for two classes 
of applications in high energy physics: at the new generation of colliders and in space 
applications. The generation rates of vacancy-interstitial pairs have been calculated as 
convolutions of the hadron spectra with the energy dependencies of the CPD. 
The time dependence of the concentrations of stable defects has been calculated for 
"standard" and "oxygenated" silicon, in conditions of continuous irradiation during 10 
years, at 20$^o$C, for generation rates of vacancy interstitial pairs corresponding to the 
applications mentioned before.
The increase of the oxygen content in irradiated silicon conduces to the increase of the 
concentrations of oxygen related defects, $VO$, $C_iO_i$ and $V_2O$, and to the decrease 
of the $V_2$, $VP$ and $C_iC_s$ ones. 
For long term operation and high generation rates of VI pairs, comparable leakage currents 
are expected in standard and oxygenated silicon p-n junctions. The beneficial influence of 
a higher oxygen content in silicon becomes visible with the decrease of the generation 
rate.

\end{document}